\documentclass[a4paper]{article}

\usepackage{INTERSPEECH2019}
\usepackage[english]{babel}

\title{Streaming Chunk-Aware Multihead Attention for Online End-to-End Speech Recognition}
\name{Shiliang Zhang$^1$, Zhifu Gao$^1$, Haoneng Luo$^2$, Ming Lei$^1$, Jie Gao$^1$, Zhijie Yan$^1$, Lei Xie$^2$}
\address{
  $^1$Speech Lab, Alibaba DAMO Academy\\\
  $^2$ASLP@NPU, Northwestern Polytechnical University, China}
\email{\{sly.zsl, zhifu.gzf\}@alibaba-inc.com, lxie@nwpu.edu.cn}

\begin{document}
\maketitle
\begin{abstract}
Recently, streaming end-to-end automatic speech recognition (E2E-ASR) has gained more and more attention. Many efforts have been paid to turn the non-streaming attention-based E2E-ASR system into streaming architecture. 
In this work, we propose a novel online E2E-ASR system by using \emph{Streaming Chunk-Aware Multihead Attention} (SCAMA) and a latency control memory equipped self-attention network (LC-SAN-M). LC-SAN-M uses chunk-level input to control the latency of encoder. As to SCAMA, a jointly trained \emph{predictor} is used to control the output of encoder when feeding to decoder, which enables decoder to generate output in streaming manner. Experimental results on the open 170-hour AISHELL-1 and an industrial-level 20000-hour Mandarin speech recognition tasks show that our approach can significantly outperform the MoChA-based baseline system under comparable setup. On the AISHELL-1 task, our proposed method achieves a character error rate (CER) of 7.39\%, to the best of our knowledge, which is the best published performance for online ASR.

\end{abstract}

\section{Introduction}
End-to-end (E2E) automatic speech recognition (ASR) has gained more and more attention in speech recognition community. Different from conventional hybrid ASR systems, E2E systems fold the acoustic model (AM), language model (LM) and pronunciation model (PM) into a single sequence to sequence model, which dramatically simplifies the training and decoding pipelines. Currently, there exists three popular end-to-end approaches, namely connectionist temporal classification (CTC)~\cite{graves2006connectionist}, recurrent neural network transducer (RNN-T)~\cite{graves2012sequence}, and attention based encoder-decoder (AED)~\cite{bahdanau2014neural,chorowski2015attention,chan2016listen}. CTC makes an independence assumption that the label outputs are conditionally independent of each other. Thereby, it usually needs to combine with an external language model in order to achieve good recognition results~\cite{graves2014towards,sak2015fast}. Unlike CTC-based model, RNN-T and attention based encoder-decoder (AED) models have no independence assumption and can achieve state-of-the-art performance even without an external language model. 

The most representative attention based model is the so-called LAS~\cite{chan2016listen}, which consists of a pyramidal bidirectional long short term memory (BLSTM) based encoder and an attention-equipped LSTM based decoder. The encoder transfers raw acoustic feature into higher-level representation, and the decoder with attention mechanism predicts the next output symbol based on the previous predictions in an auto-regressive manner. The attention module inside decoder is used to compute dynamic soft alignments and produce context vectors.
As originally defined, the soft attention needs to attend entire input sequences at each output timestep. As a result, soft attention based E2E model is inapplicable to online speech recognition, since it has to wait until the input sequence has been processed before it can generate output.

In previous works, many efforts have been made to convert full sequence soft attention into local attention, which is suitable for online speech recognition. In~\cite{raffel2017online}, a \emph{Hard Monotonic Attention} is proposed based on the insight that alignment between input and output sequence elements is monotonous in nature. Along this line, an improved monotonic attention namely \emph{Monotonic Chunkwise Attention} (MoChA) is proposed in~\cite{chiu2017monotonic}, which enables the model to perform soft attention over small chunks of the memory where a hard monotonic attention mechanism has chosen to attend. Experimental results have shown that MoChA~\cite{chiu2017monotonic} and its variants, such as AMoChA~\cite{fan2018online} and sMoChA~\cite{miao2019online},  effectively close the gap between monotonic and soft attention on speech recognition tasks. Meanwhile, works in~\cite{hou2017gaussian,tjandra2017local, merboldt2019analysis} employ local attention by computing energy values only on a local window. Both MoChA-type and local window based attentions use some preset hyper-parameters to truncate the input sequence in order to enable online attention, such as the threshold used to stop scanning memory in MoChA and the window-size in local window based attentions. These preset hyper-parameters may make these online attentions not robust enough for speech recognition in practical applications . 
Recently, works on streaming E2E-ASR try to combine attention with additional alignment information to perform streaming truncation. In~\cite{moritz2019triggered}, the proposed triggered attention (TA) uses a CTC-based classifier to dynamically control the activation of an attention-based decoder neural network. In~\cite{wang2020reducing}, the proposed Scout Network trained with the word-level force-alignment is used to streaming detect the word boundary without seeing any future frames.

Encoder architecture is another key element to streaming E2E-ASR. In order to control latency, previous works usually adopt the unidirectional LSTM~\cite{chiu2017monotonic} or latency control bidirectional LSTM (LC-BLSTM)~\cite{fan2018online,miao2019online}. More recently, self-attention based Transformer~\cite{vaswani2017attention} has become popular in E2E-ASR~\cite{dong2018speech,pham2019very}. The key improvement is the utilization
of self-attention instead of recurrent mechanism in both encoder and decoder, which enhances the abilities to capture long-range dependencies with lower computational complexity and more parallelizable training. In~\cite{tsunoo2019towards,moritz2020streaming}, Transformer is further designed into structures that are enable to perform online encoding.

In this work, we come up with a novel \emph{Streaming Chunk-Aware Multihead Attention} (SCAMA) based online E2E-ASR system. For the encoder, we extend our previous memory equipped self-attention (SAN-M)~\cite{zhifugao2020SANM} to a latency control architecture, namely LC-SAN-M. The LC-SAN-M based encoder uses chunk-level input to control the encoder latency. For SCAMA, we use a jointly trained \emph{predictor} to predict the number of tokens in each chunk and control the activation of an attention-based decoder. Compared to triggered attention (TA), the predictor is trained using a cross-entropy loss instead of CTC loss. More importantly, prediction of token number in chunk-level inputs can achieve very high accuracy, thus eliminating the mismatch between training and testing.  We have evaluated our approach on the public 170-hour AISHELL-1 and an industrial-level 20000-hour Mandarin recognition tasks. Compared to the original full sequence attention, SCAMA based online E2E-ASR system suffers from acceptable performance degradation and achieves much better performance than the baseline MoChA-based systems.

\section{E2E-ASR with Memory Equipped Self-attention}
\begin{figure}[t]
	\centering
	\includegraphics[width=0.9\linewidth]{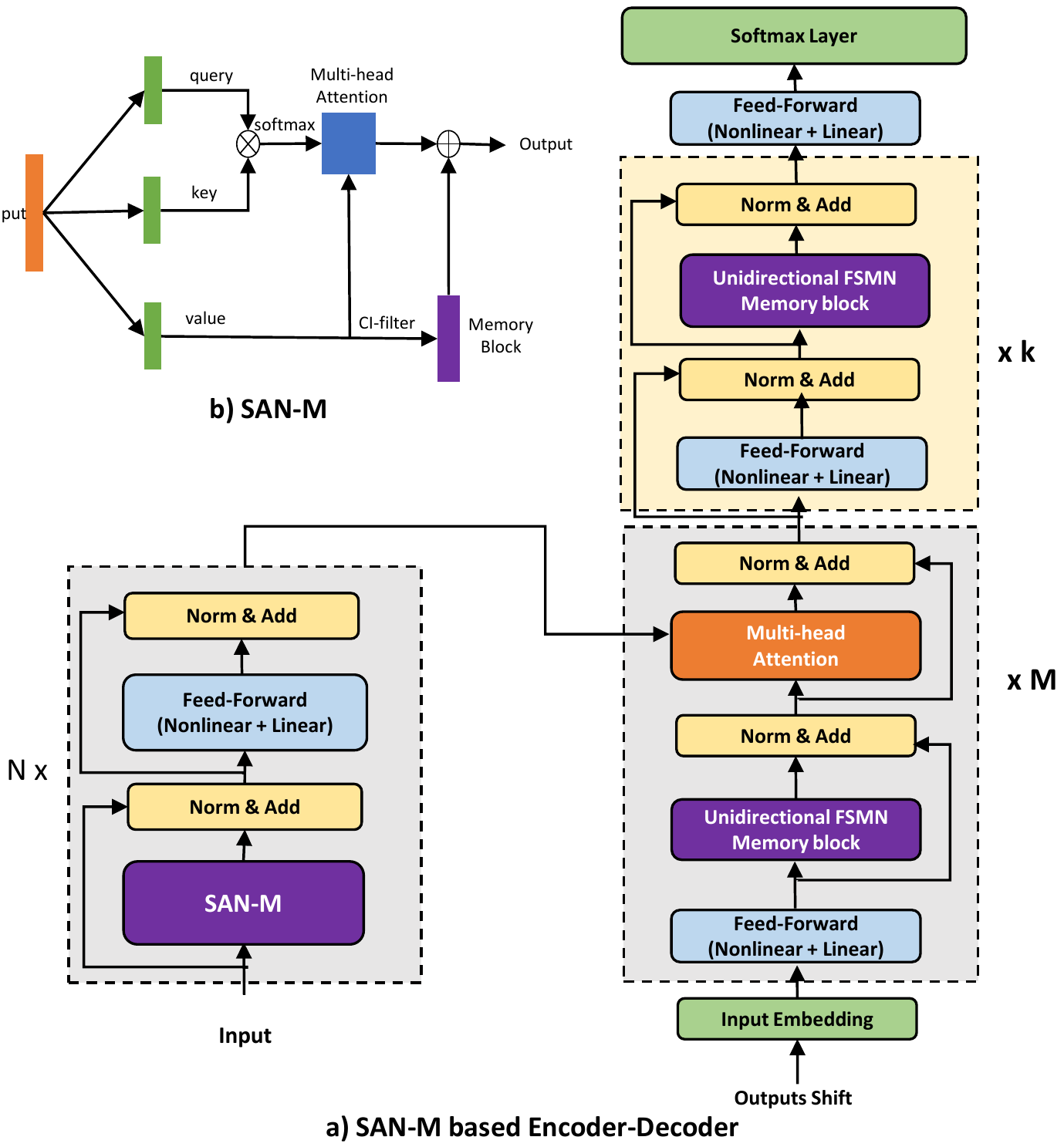}
	\caption{Illustration of memory equipped self-attention based encoder-decoder.}
	\label{fig:SAN_M}
\end{figure}
In our previous work~\cite{zhifugao2020SANM}, we have proposed a memory equipped self-attention (SAN-M) for end-to-end speech recognition in Encoder-Decoder framework. In this section, we will give a brief review on SAN-M based model. As shown in Figure~\ref{fig:SAN_M} a), the encoder consists of $N$ blocks of SAN-M and feed-forward component, the decoder consists of $M$ blocks of multihead attention equipped unidirectional deep feed-forward sequential memory network (DFSMN)~\cite{zhang2017nonrecurrent,zhang2018deep} layer, and $K$ blocks of pure unidirectional DFSMN layer. As to SAN-M, it combines the multihead self-attention~\cite{vaswani2017attention} in Transformer with the memory block in FSMN as shown in Figure~\ref{fig:SAN_M} b). This combination of context-independent FSMN memory block and context-dependent self-attention results in powerful local and long-term dependencies modeling ability. Given an input sequence, denoted as $\mathbf{X} = \{x_1, \cdots x_t, \cdots  x_T \}^T$, where each $x_t\in \mathbb{R}^{d\times 1}$ represents the input data at time instance $t$. The detailed operations of SAN-M are as follows.
\begin{equation}\label{eq.multihead_self_attention}
{\rm MultiHead}(\mathbf{X}) = [{\rm head_{1}, ..., head_{h}}] \mathbf{W}^{O}
\end{equation}
\begin{equation}\label{eq.multihead_self_attention_i}
{\rm head_{i}} = {\rm SelfAtt}(\mathbf{Q}_i, \mathbf{K}_i, \mathbf{V}_i)= {\rm softmax}(\dfrac{\mathbf{Q}_i\mathbf{K}_i^{T}}{\sqrt{d_{k}}}) \mathbf{V}_i
\end{equation}
\begin{equation}\label{eq.attention}
(\mathbf{Q}_i, \mathbf{K}_i, \mathbf{V}_i) = ( \mathbf{X}\mathbf{W}_{i}^{Q},  \mathbf{X}\mathbf{W}_{i}^{K},  \mathbf{X}\mathbf{W}_{i}^{V})
\end{equation}
Where the projections are parameter matrices $\mathbf{W}_{i}^{Q}\in \mathbb{R}^{d_{model} \times d_q}$, $\mathbf{W}_{i}^{K}\in \mathbb{R}^{d_{model}\times d_k}$, $\mathbf{W}_{i}^{V}\in \mathbb{R}^{d_{model} \times d_v}$ and $\mathbf{W}^{O}\in \mathbb{R}^{d_{model} \times hd_{v}}$. $h$ is the number of heads, and $d_{model}$ is the model dimension. In this work, the multihead attention consists of 4 heads ($h=4$). The output of FSMN memory block can be calculated as follows.
\begin{equation}\label{eq.fsmn_1}
m_t = v_t +\sum\limits_{i = 0}^{L_\ell -1 } a_i \odot v_{t-i} + \sum\limits_{j = 1}^{L_r } c_i \odot v_{t+j} \\
\end{equation}
\begin{equation}\label{eq.fsmn_2}
{\bf M} = [m_1, m_2, \cdots, m_T]^T
\end{equation}
Here, $v_t$ denotes the $t$-th time instance in self-attention values. $L_\ell$ and $L_r$ are the look-back and look-ahead order of FSMN memory block respectively. $\odot$ denotes the element-wise multiplication of two equally-sized vectors.
\begin{equation}\label{eq.output}
\mathbf{Y} = {\rm MultiHead}(\mathbf{X}) + {\bf M}
\end{equation}
$\mathbf{Y}$ denotes the output of the SAN-M. The other operations \emph{Feed-Forward}, \emph{Norm $\&$ Add} and \emph{Multi-head Attention} are the same to original Transformer~\cite{vaswani2017attention} and the \emph{unidirectional FSMN} is the same to original FSMN~\cite{zhang2017nonrecurrent}.
\section{Online E2E-ASR}
\begin{figure}[t]
	\centering
	\includegraphics[width=0.9\linewidth]{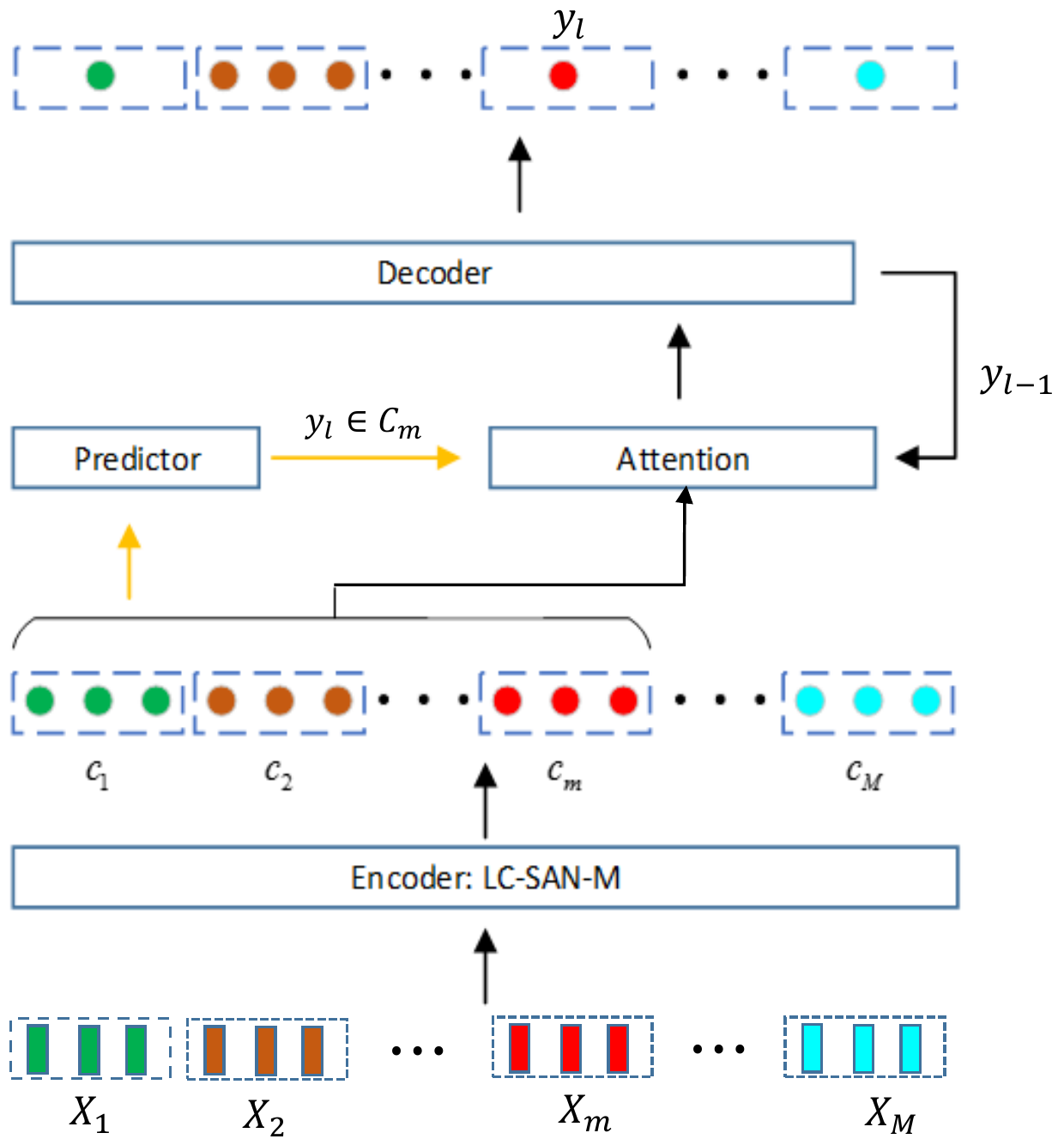}
	\caption{Illustration of SCAMA based online E2E-ASR system.}
	\label{fig:scama_overall}
\end{figure}
The overall architecture of the proposed online end-to-end speech recognition system is as shown in Figure~\ref{fig:scama_overall}. Compared to the original SAN-M based E2E model, there are two changes to make the system streamable. As to the encoder, we extend SAN-M to a latency control version, namely LC-SAN-M. As to the attention module, we come up with a novel Streaming Chunk-Aware Multihead Attention. 
\subsection{LC-SAN-M}
For original memory equipped self-attention (SAN-M) based encoder in~\cite{zhifugao2020SANM}, the full sequence attention mechanism makes it unsuitable for online ASR. In order to control the encoder latency, we extend the SAN-M to LC-SAN-M. The input sequence $\mathbf{X}$ is divided into chunk-level according to a preset chunk size $c$, denoted as $\mathbf{X} = \{[x_1, \cdots, x_c],[x_{c+1}, \cdots, x_{2c}],\cdots,[x_{nc+1} \cdots x_{T}] \}^T$. The chunk-size $c$ is related to the encoder latency. In this work, we will evaluate various chunk-sizes with value being 5, 10 and 15.  Notationally, $\mathbf{X}_k=\{[x_{kc+1}, \cdots, x_{(k+1)c}]\}^T$ denotes the samples in $k$-th chunk. For each time instance in $k$-th chunk, it can only access samples in the current chunk and previous chunks. Thereby, the output of LC-SAN-M for $\mathbf{X}_k$ can be calculated using the following formulations.
\begin{equation}\label{eq.lc_san_m_1}
(\mathbf{Q}_i(k), \mathbf{K}_i(k), \mathbf{V}_i(k)) = (\mathbf{X}_k\mathbf{W}_{i}^{Q}, \mathbf{X}_k\mathbf{W}_{i}^{K}, \mathbf{X}_k\mathbf{W}_{i}^{V})
\end{equation}
\begin{equation}\label{eq.lc_san_m_2}
\bar{\mathbf{K}}_i(k) = [\bar{\mathbf{K}}_i(k-1); \mathbf{K}_i(k)]
\end{equation}
\begin{equation}\label{eq.lc_san_m_3}
\bar{\mathbf{V}}_i(k) = [\bar{\mathbf{V}}_i(k-1); \mathbf{V}_i(k)]
\end{equation}
\begin{equation}\label{eq.multihead_self_attention_i}
{\rm head_{i}}(k) = {\rm SelfAtt}(\mathbf{Q}_i(k), \bar{\mathbf{K}}_i(k), \bar{\mathbf{V}}_i(k))
\end{equation}
\begin{equation}\label{eq.multihead_self_attention}
{\rm MultiHead}(\mathbf{X}_k) =  [{\rm head_{1}(k), ..., head_{h}(k)}]\mathbf{W}^{O}
\end{equation}
Furthermore, Eq. (\ref{eq.fsmn_1}) is modified to the following unidirectional FSMN memory block.
\begin{equation}
m_t = v_t +\sum\limits_{i = 0}^{L -1 } a_i \odot v_{t-i}, t\in [kc+1,\cdots,(k+1)c] \\
\end{equation}
\begin{equation}\label{eq.fsmn_2}
{\bf M}_k = [m_{kc+1}, \cdots, m_{(k+1)c}]^T
\end{equation}
Here, $L$ is the total filter orders of FSMN memory block. Finally, we can get the output of LC-SAN-M for $\mathbf{X}_k$ as follow.
\begin{equation}\label{eq.output}
\mathbf{Y}_k = {\rm MultiHead}(\mathbf{X}_k) + {\bf M}_k
\end{equation}
\subsection{SCAMA}
As shown in Figure~\ref{fig:scama_overall}, we stack a \emph{predictor} on the top of encoder, which is trained to predict the number of tokens in each chunk. The chunked outputs of encoder are spliced and then fed into the predictor. Notationally, let's denote the spliced output of $k$-th chunk as $\mathbf{h}_k^s$. Then the predictor generate the probability $p_k$ as :
\begin{equation}\label{eq.predictor_output}
p_k = {\rm softmax}(\max(\mathbf{h}_k^s\mathbf{W}^1 + \mathbf{b^1},0)\mathbf{W}^2 + \mathbf{b^2})
\end{equation}
The predictor is trained using the cross-entropy loss:
\begin{equation}\label{eq.predictor_ce_loss}
\mathcal{L}_{pred} = -\sum_{k} t_k log (p_k)
\end{equation}
And the overall system is jointly optimized using the following loss function:
\begin{equation}\label{eq.scama_loss}
\mathcal{L} =\mathcal{L}_{e2e} + \alpha \mathcal{L}_{pred}
\end{equation}
Here, $t_k$ denotes the one-hot vector of the ground truth token number in $k$-th chunk and $\alpha$ is 0.2. $\mathcal{L}_{e2e}$ is the original CE-loss to train the encoder-decoder. We use a well-trained CTC-based ASR system~\cite{zhang2019investigation} to generate the frame-level alignments and then convert them into the chunk-level labels. We first count the maximum number of tokens contained in the chunk from training set. Thereby, training of predictor is formulated as a multi-class classification problem. During training, the ground truth token number is used to guide the encoder output fed into the decoder. If the $\ell$-th token of decoder is in $m$-th chunk, then only $c_1$ to $c_m$ chunks are
fed into the attention module to generate the context vector for decoder. During inference, the class with the maximum probability is chose as the output for predictor, which is used to guide how many steps the decoder should attend to the current input chunk.
\subsection{Decoding Strategy}
For encoder-decoder based E2E-ASR, the inference is terminated when an end-of-sentence ($<$eos$>$) token is predicted. For streaming E2E-ASR, one of the issues is that the decoder may predict the $<$eos$>$ token too early or too late~\cite{li2020towards}. In our works, we also find that the decoder may generate the $<$eos$>$ token too early, especially when the chunk size is small. We propose a trick during beam search based decoding to handle this problem. During inference, if the decoder generate an $<$eos$>$ token with the input is not the last chunk, we will use the previous token and historical information to predict the next token instead of $<$eos$>$. As to the last chunk, if the predicted token number is $N$, the total decoding steps will be $1$ to $N+2$. The inference is terminated when the decoder generated an $<$eos$>$ token or decoded for $N+2$ steps in the last chunk. 
\newcommand{\pp}[1]{\raisebox{-1.2ex}[0pt][0pt]{\shortstack{#1}}}
\begin{table}[t]
	\centering
	\begin{footnotesize}
		\caption{Performance of various E2E models on AISHELL-1. (FSA denotes the full sequence attention.)}
		\begin{tabular}[t]{|c|c|c|c|c|}
			\hline
			EXP &Encoder          & Decoder      & Attention & CER(\%)\\ \hline
			1  & SAN-M(10)        & DFSMN(3)     &  FSA      & 6.46 \\ \hline
			2  &\pp{LC-SAN-M(10)} &\pp{DFSMN(3)} &  FSA      & 6.92 \\\cline{4-5}
			3  &                  &              &  SCAMA    & 7.39 \\\hline 
			4  &\pp{LC-SAN-M(10)} & \pp{LSTM(3)} &  FSA      & 8.78 \\\cline{4-5}
			5  &                  &              &  MoChA    & 9.01 \\\hline 
		\end{tabular}
		\label{tab:AISHELL_res1}
	\end{footnotesize}
\end{table}
\begin{table}[t]
	\centering
		\begin{footnotesize}
	\caption{Comparison of systems on AISHELL-1 task. ``Y'' denotes ``Yes'' and ``N'' denotes ``No''.}
	\begin{tabular}[t]{|c|c|c|c|c|}
		\hline
		Model                                      & E2E    & LM & Online & CER(\%)  \\\hline
		LAS\cite{shan2019component}                & Y      & Y  &  N     & 8.71 \\\hline
		CTC\&attention\cite{karita2019comparative} & Y      & Y  &  N     & {6.70} \\\hline
		TDNN-LFMMI\cite{bu2017aishell}             & N      & Y  &  Y     &  7.62 \\\hline
		Transformer-MoChA\cite{tsunoo2019towards}  & Y      & Y  &  Y     & 9.7 \\\hline
		\bf{LC-SAN-M-SCAMA}                        & Y      & N  &  Y     & \bf{7.39} \\\hline 	               
	\end{tabular}
	\label{tab:AISHLL_state_of_art}
\end{footnotesize}
\end{table}
\begin{table*}[t]
	\centering
	\begin{scriptsize}
		\caption{Performance of various E2E models on 20000 hours task .}
		\begin{tabular}[t]{|c|c|c|c|c|c|c|c|c|c|c|}
			\hline
			Model Type      &\multicolumn{2}{c|}{CTC} &\multicolumn{4}{c|}{Non-streaming E2E}                & \multicolumn{4}{c|}{Streaming E2E} \\\hline
			Model ID        &  CTC1    & CTC2    & E2E1       &E2E2        &E2E3              & E2E4         & E2E5        & E2E6        & E2E7  & E2E8 \\\hline
			Encoder         &DFSMN(10) & DFSMN(20)& SAN-M(40) &\multicolumn{7}{c|}{LC-SAN-M(40)} \\\hline
			Decoder         &  -       &  -      & DFSMN(12)  &\multicolumn{2}{c|}{DFSMN(12)} & \multicolumn{2}{c|}{LSTM(4)} & \multicolumn{3}{c|}{DFSMN(12)} \\\hline
			Attention       &  -       &  -      & FSA        &FSA         &FSA               & FSA          & MoChA       & \multicolumn{3}{c|}{SCAMA} \\\hline
			Encoder Latency &  600ms   & 1.2s    & Full       &600ms       &900ms             & 600ms        & 600ms       & 300ms       & 600ms & 900ms\\\hline
			Total Latency   &  600ms   & 1.2s    & Full       &Full        &Full              & Full         & 600ms       & 300ms       & 600ms & 900ms \\\hline
			Parameter(M)    &  25      & 45      & 42         &42          &42                & 60           & 60          & 43          & 43    &  43\\\hline
			Common Set (CER\%)& 11.6   & 9.9     & 9.0        &9.9         &9.4               & 10.1         & 11.5        & 11.4        & 10.4   &  9.8 \\\hline
			Far-Field Set (CER\%)& 20.3& 17.7    & 13.7       &15.1        &14.9              & 15.8         & 18.1        & 17.0        & 16.0  &  15.2\\\hline
		\end{tabular}
		\label{tab:fly_2W_Res}
	\end{scriptsize}
\end{table*}
\section{Experiments} 
We have evaluated our proposed SCAMA based online E2E-ASR on two Mandarin speech recognition tasks, namely the AISHELL-1 task released in~\cite{bu2017aishell} and a 20000-hour Mandarin task. For AISHELL-1 task, we use the 150-hour training set for model training and use the 10-hour development set for early-stopping. Finally, the character error rate (CER\%) is reported in the 7176-sentence (about 5 hours) test set.  For the 20000-hour Mandarin task, the same as in~\cite{zhang2019investigation},  which consists of about 20000 hours data that collected from multi-domain, including news,
sport, tourism, game, literature, education et al. It is divided into training set and development set according to the ratio of 95\% to 5\%. A \emph{far-field set} consists of about 15 hours data and a \emph{common set} consists of about 30 hours data are used to evaluated the performance.

Acoustic feature used for all experiments are 80-dimensional log-mel filter-bank (FBK) energies computed on 25ms window with 10ms shift. We stack the consecutive frames within a context window of 7 (3+1+3) to produce the 560-dimensional features and then down-sample the inputs frame rate to 60ms. Acoustic modeling units are the Chinese characters, which are 4233 and 9000 for AISHELL-1 and 20000-hour tasks respectively. We use the Tensorflow~\cite{abadi2016tensorflow} to train the model in a distributed manner. 
Label smoothing and dropout regularization with value being 0.1 are added to prevent over-fitting. SpecAugment~\cite{park2019specaugment} is also used in all experiments.
\subsection{AISHELL-1 Task}
\label{sec:aishell}
In Table~\ref{tab:AISHELL_res1}, we have evaluated the performance of various E2E-ASR system on AISHELL-1 task. For the baseline models, we trained the SAN-M based E2E-ASR system~\cite{zhifugao2020SANM}. As shown in Figure~\ref{fig:SAN_M}, we set $N=10$, $M=3$ and $K=0$, which means the encoder consists of 10 SAN-M blocks and the decoder consists of three full sequence attention (FSA) equipped DFSMN layers. The linear and nonlinear layer size is 512 and 2048 in both encoder and decoder respectively. This model achieve a CER of 6.46\%. We then replace the SAN-M based encoder with the LC-SAN-M based encoder to investigate the influence of encoder latency to the performance. The chunk-size of LC-SAN-M is 10. Comparison of \emph{exp1} and \emph{exp2} in Table~\ref{tab:AISHELL_res1} shown that it suffers from about 7\% relative performance degradation. Furthermore, we replace the full sequence attention (FSA) with the proposed SCAMA. As shown in \emph{exp3} of Table~\ref{tab:AISHELL_res1}, it suffers from 6.8\% relative performance degradation. In order to evaluate the influence of decoder architecture and compare with the MoChA, we have conducted \emph{exp4} and \emph{exp5}. Both decoder consists of three LSTM layer with 512 units. Experimental results show that replace the full sequence attention with MoChA based online attention suffers from small performance degradation in this task. However, the performance of LSTM-decoder based systems are far behind the DFSMN-decoder based systems. In Table~\ref{tab:AISHLL_state_of_art}, we have compared our proposed system with the other published systems on this task. Our proposed LC-SAN-M with SCAMA based online E2E-ASR system achieve a CER of 7.39\% without using any external LM. To our best knowledge, this is the state-of-the-art performance for online ASR system in this task. 

\subsection{20000-hour Task}
In this task, we have compared three types of E2E-ASR systems: \emph{CTC}, \emph{non-streaming E2E} and \emph{streaming E2E}. For CTC-based systems, as in~\cite{zhang2019investigation}, we have trained two DFSMN-CTC-sMBR systems with 10 and 20 DFSMN-layers, denoted as \emph{CTC1} and \emph{CTC2} in Table~\ref{tab:fly_2W_Res} respectively. CTC-based models are decoded with an external 5-gram language model. For non-streaming E2E-ASR systems, we have trained four models, denoted as \emph{E2E1} to \emph{E2E4} in Table~\ref{tab:fly_2W_Res}. For \emph{E2E1} system, we set the $N=40$, $M=6$ and $K=6$, which means the decoder consists of 12 DFSMN-layers with the bottom 6 layers equipped with full sequence multihead attention. The linear and nonlinear layer size is 256 and 1024 respectively. For \emph{E2E2} to \emph{E2E4}, we replace the sequence-level SAN-M based encoder with the LC-SAN-M based encoder. The encoder chunk-size is 10 for \emph{E2E2} and \emph{E2E4} and is 15 for \emph{E2E3}. Since we down-sample the input with 6, the corresponding encoder latency is 600ms and 900ms respectively. For \emph{E2E4}, we further replace the decoder with LSTM, which consists of 4 LSTM layers with 768 units. For streaming E2E-ASR system, we have evaluated MoChA based model (\emph{E2E5}) and SCAMA based models with chunk size being 5, 10 and 15 (\emph{E2E6} to \emph{E2E8}). It takes about 3 days to train a SCAMA model when using 32 NVIDIA TESLA V100 GPUs, which is 3 times faster than the MoChA system. 

In Table~\ref{tab:fly_2W_Res}, we have summarized the performance of various systems on 20000-hour task. For non-streaming E2E-ASR systems, when the encoder's future contextual information is limited, performance will degrade. Compared \emph{E2E2} with \emph{E2E4}, the DFSMN-decoder based system achieve better performance than the LSTM-decoder based system as well as smaller in model size. For \emph{E2E4} and \emph{E2E5}, when the full sequence attention is replaced with online MoChA attention, performance will suffer from significant loss. This experimental phenomena is different to the AISHELL-1 task in Sec.~\ref{sec:aishell}. According to our experimental analysis, this is due to MoChA use a preset threshold to stop scanning memory during inference is not robust to noisy speech.  Comparison of SCAMA based systems (\emph{E2E6} to \emph{E2E8}) with the non-streaming systems (\emph{E2E1} to \emph{E2E3}) shows that SCAMA based online attention suffer from acceptable performance degradation. Figure~\ref{fig:scama_overall} is the visualization of attention in the last layer of \emph{E2E1} and \emph{E2E7}. The general trend of full sequence multihead attention is monotonous. However, it seems that attention between the encoder and decoder not only plays the role of alignment but also conducts context modeling. Thereby, restrict attention to local window or completely monotonous will suffer from performance degradation. As to SCAMA based systems, the performance degradation is less than MoChA based system since it only limits the future information.
\begin{figure}[t]
	\centering
	\includegraphics[width=0.9\linewidth]{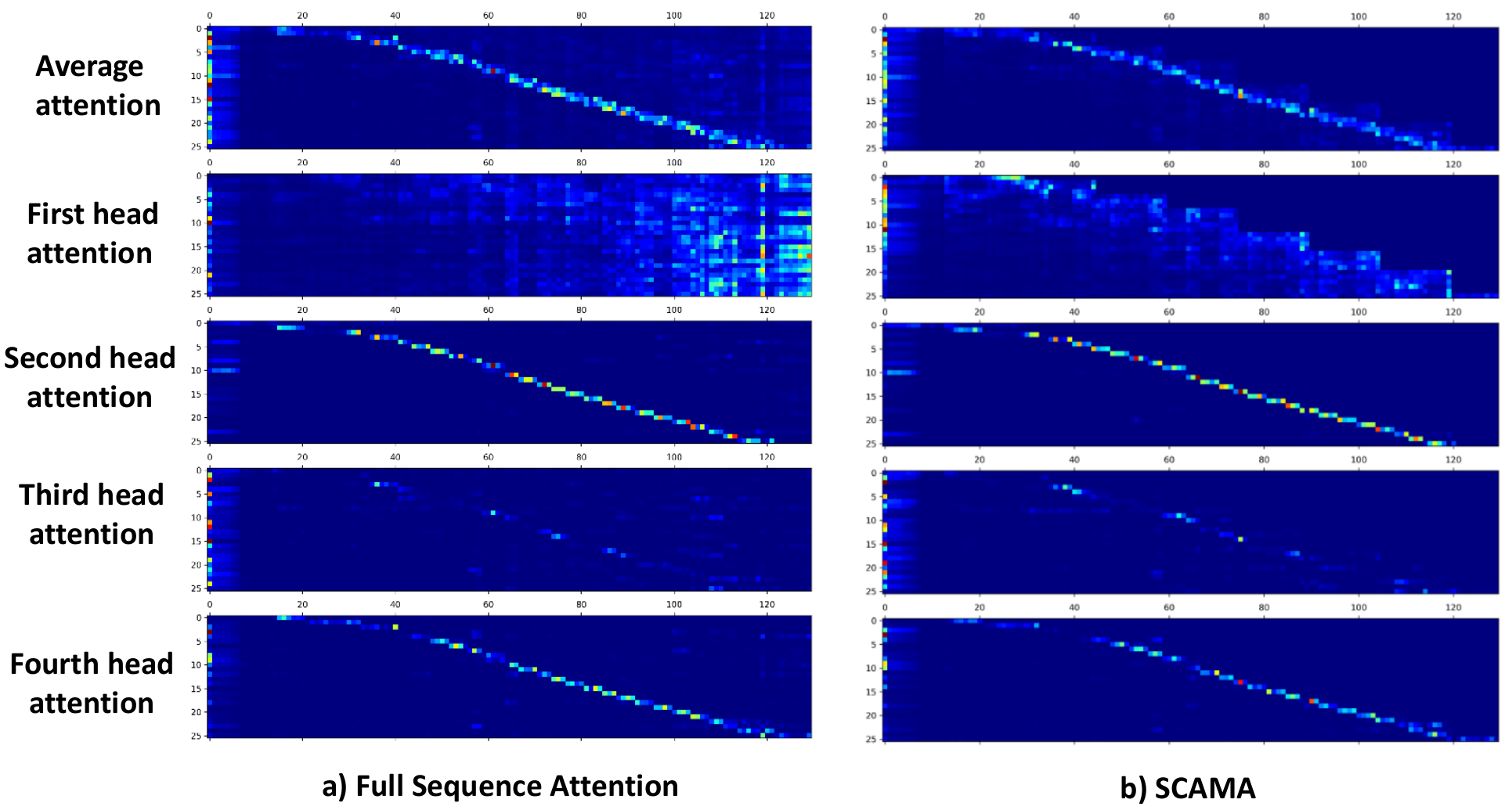}
	\caption{Illustration of FSA and SCAMA. x-axis refer to acoustic frames and y-axis refer to characters.}
	\label{fig:scama_overall}
\end{figure}
\section{Conclusions}
In this paper, we have proposed a novel online end-to-end speech recognition system. Specially, we have come up with a \emph{Streaming Chunk-Aware Multihead Attention} (SCAMA) and a latency control memory equipped self-attention (LC-SAN-M) based online E2E-ASR system. Compared to full sequence attention, the performance degradation of SCAMA is acceptable. On AISHELL-1 task, our proposed online E2E-ASR system achieves a CER of 7.39\% without using any external LM. On a 20000-hour Mandarin task,  SCAMA based online E2E-ASR system can significantly outperform the CTC and MoChA based systems with the same latency.

\bibliographystyle{IEEEtran}

\bibliography{mybib}

\begin{thebibliography}{10}
\providecommand{\url}[1]{#1}
\csname url@samestyle\endcsname
\providecommand{\newblock}{\relax}
\providecommand{\bibinfo}[2]{#2}
\providecommand{\BIBentrySTDinterwordspacing}{\spaceskip=0pt\relax}
\providecommand{\BIBentryALTinterwordstretchfactor}{4}
\providecommand{\BIBentryALTinterwordspacing}{\spaceskip=\fontdimen2\font plus
\BIBentryALTinterwordstretchfactor\fontdimen3\font minus
  \fontdimen4\font\relax}
\providecommand{\BIBforeignlanguage}[2]{{%
\expandafter\ifx\csname l@#1\endcsname\relax
\typeout{** WARNING: IEEEtran.bst: No hyphenation pattern has been}%
\typeout{** loaded for the language `#1'. Using the pattern for}%
\typeout{** the default language instead.}%
\else
\language=\csname l@#1\endcsname
\fi
#2}}
\providecommand{\BIBdecl}{\relax}
\BIBdecl

\bibitem{graves2006connectionist}
A.~Graves, S.~Fern{\'a}ndez, F.~Gomez, and J.~Schmidhuber, ``Connectionist
  temporal classification: labelling unsegmented sequence data with recurrent
  neural networks,'' in \emph{Proceedings of the 23rd international conference
  on Machine learning}, 2006, pp. 369--376.

\bibitem{graves2012sequence}
A.~Graves, ``Sequence transduction with recurrent neural networks,''
  \emph{arXiv preprint arXiv:1211.3711}, 2012.

\bibitem{bahdanau2014neural}
D.~Bahdanau, K.~Cho, and Y.~Bengio, ``Neural machine translation by jointly
  learning to align and translate,'' \emph{arXiv preprint arXiv:1409.0473},
  2014.

\bibitem{chorowski2015attention}
J.~K. Chorowski, D.~Bahdanau, D.~Serdyuk, K.~Cho, and Y.~Bengio,
  ``Attention-based models for speech recognition,'' in \emph{Advances in
  neural information processing systems}, 2015, pp. 577--585.

\bibitem{chan2016listen}
W.~Chan, N.~Jaitly, Q.~Le, and O.~Vinyals, ``Listen, attend and spell: A neural
  network for large vocabulary conversational speech recognition,'' in
  \emph{2016 IEEE International Conference on Acoustics, Speech and Signal
  Processing (ICASSP)}.\hskip 1em plus 0.5em minus 0.4em\relax IEEE, 2016, pp.
  4960--4964.

\bibitem{graves2014towards}
A.~Graves and N.~Jaitly, ``Towards end-to-end speech recognition with recurrent
  neural networks,'' in \emph{International conference on machine learning},
  2014, pp. 1764--1772.

\bibitem{sak2015fast}
H.~Sak, A.~Senior, K.~Rao, and F.~Beaufays, ``Fast and accurate recurrent
  neural network acoustic models for speech recognition,'' \emph{arXiv preprint
  arXiv:1507.06947}, 2015.

\bibitem{raffel2017online}
C.~Raffel, M.-T. Luong, P.~J. Liu, R.~J. Weiss, and D.~Eck, ``Online and
  linear-time attention by enforcing monotonic alignments,'' in
  \emph{Proceedings of the 34th International Conference on Machine
  Learning-Volume 70}.\hskip 1em plus 0.5em minus 0.4em\relax JMLR. org, 2017,
  pp. 2837--2846.

\bibitem{chiu2017monotonic}
C.-C. Chiu and C.~Raffel, ``Monotonic chunkwise attention,'' \emph{arXiv
  preprint arXiv:1712.05382}, 2017.

\bibitem{fan2018online}
R.~Fan, P.~Zhou, W.~Chen, J.~Jia, and G.~Liu, ``An online attention-based model
  for speech recognition,'' \emph{arXiv preprint arXiv:1811.05247}, 2018.

\bibitem{miao2019online}
H.~Miao, G.~Cheng, P.~Zhang, T.~Li, and Y.~Yan, ``Online hybrid ctc/attention
  architecture for end-to-end speech recognition,'' \emph{Proc. of Interspeech
  2019}, pp. 2623--2627, 2019.

\bibitem{hou2017gaussian}
J.~Hou, S.~Zhang, and L.-R. Dai, ``Gaussian prediction based attention for
  online end-to-end speech recognition.'' in \emph{INTERSPEECH}, 2017, pp.
  3692--3696.

\bibitem{tjandra2017local}
A.~Tjandra, S.~Sakti, and S.~Nakamura, ``Local monotonic attention mechanism
  for end-to-end speech and language processing,'' \emph{arXiv preprint
  arXiv:1705.08091}, 2017.

\bibitem{merboldt2019analysis}
A.~Merboldt, A.~Zeyer, R.~Schl{\"u}ter, and H.~Ney, ``An analysis of local
  monotonic attention variants,'' \emph{Proc. of Interspeech 2019}, pp.
  1398--1402, 2019.

\bibitem{moritz2019triggered}
N.~Moritz, T.~Hori, and J.~Le~Roux, ``Triggered attention for end-to-end speech
  recognition,'' in \emph{ICASSP 2019-2019 IEEE International Conference on
  Acoustics, Speech and Signal Processing (ICASSP)}.\hskip 1em plus 0.5em minus
  0.4em\relax IEEE, 2019, pp. 5666--5670.

\bibitem{wang2020reducing}
C.~Wang, Y.~Wu, S.~Liu, J.~Li, L.~Lu, G.~Ye, and M.~Zhou, ``Reducing the
  latency of end-to-end streaming speech recognition models with a scout
  network,'' \emph{arXiv preprint arXiv:2003.10369}, 2020.

\bibitem{vaswani2017attention}
A.~Vaswani, N.~Shazeer, N.~Parmar, J.~Uszkoreit, L.~Jones, A.~N. Gomez,
  {\L}.~Kaiser, and I.~Polosukhin, ``Attention is all you need,'' in
  \emph{Advances in neural information processing systems}, 2017, pp.
  5998--6008.

\bibitem{dong2018speech}
L.~Dong, S.~Xu, and B.~Xu, ``Speech-transformer: a no-recurrence
  sequence-to-sequence model for speech recognition,'' in \emph{2018 IEEE
  International Conference on Acoustics, Speech and Signal Processing
  (ICASSP)}.\hskip 1em plus 0.5em minus 0.4em\relax IEEE, 2018, pp. 5884--5888.

\bibitem{pham2019very}
N.-Q. Pham, T.-S. Nguyen, J.~Niehues, M.~Muller, and A.~Waibel, ``Very deep
  self-attention networks for end-to-end speech recognition,'' \emph{arXiv
  preprint arXiv:1904.13377}, 2019.

\bibitem{tsunoo2019towards}
E.~Tsunoo, Y.~Kashiwagi, T.~Kumakura, and S.~Watanabe, ``Towards online
  end-to-end transformer automatic speech recognition,'' \emph{arXiv preprint
  arXiv:1910.11871}, 2019.

\bibitem{moritz2020streaming}
N.~Moritz, T.~Hori, and J.~L. Roux, ``Streaming automatic speech recognition
  with the transformer model,'' \emph{arXiv preprint arXiv:2001.02674}, 2020.

\bibitem{zhifugao2020SANM}
Z.~Gao, S.~Zhang, and M.~Lei, ``{SAN-M}: Memory equipped self-attention for
  end-to-end speech recognition,'' in \emph{Submitted to INTERSPEECH 2020}.

\bibitem{zhang2019investigation}
S.~Zhang, M.~Lei, Y.~Liu, and W.~Li, ``Investigation of modeling units for
  mandarin speech recognition using {DFSMN-CTC-sMBR},'' in \emph{2019 IEEE
  International Conference on Acoustics, Speech and Signal Processing
  (ICASSP)}.\hskip 1em plus 0.5em minus 0.4em\relax IEEE, 2019, pp. 7085--7089.

\bibitem{zhang2017nonrecurrent}
S.~Zhang, C.~Liu, H.~Jiang, S.~Wei, L.~Dai, and Y.~Hu, ``Nonrecurrent neural
  structure for long-term dependence,'' \emph{IEEE/ACM Transactions on Audio,
  Speech, and Language Processing}, vol.~25, no.~4, pp. 871--884, 2017.

\bibitem{zhang2018deep}
S.~Zhang, M.~Lei, Z.~Yan, and L.~Dai, ``{Deep-FSMN} for large vocabulary
  continuous speech recognition,'' in \emph{2018 IEEE International Conference
  on Acoustics, Speech and Signal Processing (ICASSP)}.\hskip 1em plus 0.5em
  minus 0.4em\relax IEEE, 2018, pp. 5869--5873.

\bibitem{li2020towards}
B.~Li, S.-y. Chang, T.~N. Sainath, R.~Pang, Y.~He, T.~Strohman, and Y.~Wu,
  ``Towards fast and accurate streaming end-to-end asr,'' in \emph{ICASSP
  2020-2020 IEEE International Conference on Acoustics, Speech and Signal
  Processing (ICASSP)}.\hskip 1em plus 0.5em minus 0.4em\relax IEEE, 2020, pp.
  6069--6073.

\bibitem{shan2019component}
C.~Shan, C.~Weng, G.~Wang, D.~Su, M.~Luo, D.~Yu, and L.~Xie, ``Component
  fusion: Learning replaceable language model component for end-to-end speech
  recognition system,'' in \emph{ICASSP 2019-2019 IEEE International Conference
  on Acoustics, Speech and Signal Processing (ICASSP)}.\hskip 1em plus 0.5em
  minus 0.4em\relax IEEE, 2019, pp. 5361--5635.

\bibitem{karita2019comparative}
S.~Karita, N.~Chen, T.~Hayashi, T.~Hori, H.~Inaguma, Z.~Jiang, M.~Someki,
  N.~E.~Y. Soplin, R.~Yamamoto, X.~Wang \emph{et~al.}, ``A comparative study on
  transformer vs {RNN} in speech applications,'' \emph{arXiv preprint
  arXiv:1909.06317}, 2019.

\bibitem{bu2017aishell}
H.~Bu, J.~Du, X.~Na, B.~Wu, and H.~Zheng, ``{AISHELL}-1: An open-source
  mandarin speech corpus and a speech recognition baseline,'' in \emph{2017
  20th Conference of the Oriental Chapter of the International Coordinating
  Committee on Speech Databases and Speech I/O Systems and Assessment
  (O-COCOSDA)}.\hskip 1em plus 0.5em minus 0.4em\relax IEEE, 2017, pp. 1--5.

\bibitem{abadi2016tensorflow}
M.~Abadi, P.~Barham, J.~Chen, Z.~Chen, A.~Davis, J.~Dean, M.~Devin,
  S.~Ghemawat, G.~Irving, M.~Isard \emph{et~al.}, ``Tensorflow: A system for
  large-scale machine learning,'' in \emph{12th $\{$USENIX$\}$ Symposium on
  Operating Systems Design and Implementation ($\{$OSDI$\}$ 16)}, 2016, pp.
  265--283.

\bibitem{park2019specaugment}
D.~S. Park, W.~Chan, Y.~Zhang, C.-C. Chiu, B.~Zoph, E.~D. Cubuk, and Q.~V. Le,
  ``Specaugment: A simple data augmentation method for automatic speech
  recognition,'' \emph{arXiv preprint arXiv:1904.08779}, 2019.

\end{thebibliography}

\end{document}